\documentclass[10pt,aps,twocolumn,secnumarabic,balancelastpage,amsmath,amssymb,nofootinbib,hyperref=pdftex,prl,citeautoscript,showpacs]{revtex4-1}

\usepackage{graphics}
\usepackage[pdftex]{graphicx}
\usepackage{tabularx}
\graphicspath{{./Figures/}}
\usepackage{longtable}
\usepackage{amsmath}
\usepackage{color,soul}
\usepackage{braket}

\makeatletter
\def\p@subsection{\thesection\,}
\makeatother

\newcommand{\bo}[1]{{\bf #1}}

\newcommand{\x}{\bo{r}}

\newcommand{\od}{\text{1D}}
\newcommand{\xp}{{\bf r}_\perp}

\usepackage{hyperref}
\hypersetup{colorlinks=true,linkcolor=blue,anchorcolor=blue,citecolor=blue,filecolor=blue,urlcolor=blue,bookmarksnumbered=true,pdfview=FitB}

\begin{document}

\title{DIII Topological Superconductivity with Emergent Time-Reversal Symmetry}
\author{Christopher Reeg}
\author{Constantin Schrade}
\author{Jelena Klinovaja}
\author{Daniel Loss}
\affiliation{Department of Physics, University of Basel, Klingelbergstrasse 82, CH-4056 Basel, Switzerland}
\date{\today}
\begin{abstract}
We find a new class of topological superconductors which possess an emergent time-reversal symmetry that is present only after projecting to an effective low-dimensional model. We show that a topological phase in symmetry class DIII can be realized in a noninteracting system coupled to an $s$-wave superconductor only if the physical time-reversal symmetry of the system is broken, and we provide three general criteria that must be satisfied in order to have such a phase. We also provide an explicit  model which realizes the class DIII topological superconductor in 1D. We show that, just as in time-reversal invariant topological superconductors, the topological phase is characterized by a Kramers pair of Majorana fermions that are protected by the emergent time-reversal symmetry.
\end{abstract}

\maketitle

\paragraph*{Introduction.}  Topological superconductors have been intensively pursued in recent years \cite{Alicea:2012,Leijnse:2012,Beenakker:2013} because the Majorana fermions which are localized to their boundaries have potential applications in the development of a topological quantum computer \cite{Kitaev:2001,Nayak:2008}. The most promising proposals to date for engineering topological superconductivity involve coupling a conventional superconductor either to a nanowire with Rashba spin-orbit interaction that is subjected to an external magnetic field \cite{Lutchyn:2010,Oreg:2010,Mourik:2012,Deng:2012,Das:2012,Rokhinson:2012,Churchill:2013,Finck:2013,Chang:2015,Albrecht:2016,Deng:2016,Zhang:2017} 
or to a ferromagnetic atomic chain \cite{NadjPerge:2013,Pientka:2013,Klinovaja:2013,Vazifeh:2013,Braunecker:2013,NadjPerge:2014,Peng:2015,Pawlak:2016}.  

Additionally, there have been several proposals to engineer topological superconductors in symmetry class DIII. Such systems possess both particle-hole symmetry and time-reversal symmetry \cite{Schnyder:2008}, with the presence of time-reversal symmetry ensuring that the Majorana fermions existing at the boundaries of class DIII topological superconductors come in Kramers pairs. In one dimension (1D), where superconductivity is required to be induced by the proximity effect, it has been shown that a nontrivial topological phase in class DIII can be realized by proximity coupling a noninteracting multichannel Rashba nanowire to an unconventional superconductor \cite{Wong:2012,Nakosai:2013,Zhang:2013,Dumitrescu:2014} or to two conventional superconductors forming a Josephson junction with a phase difference of $\pi$ \cite{Keselman:2013}. Alternatively, an effective $\pi$-phase difference can be induced in a multichannel Rashba nanowire with repuslive electron-electron interactions \cite{Haim:2014} or in a system of two topological insulators coupled to a conventional superconductor via a magnetic insulator \cite{Schrade:2015}. It has also been proposed to realize class DIII topological superconductivity in a system of two Rashba nanowires \cite{Klinovaja:2014,Gaidamauskas:2014,Schrade:2017} or two topological insulators \cite{Klinovaja:2014b} coupled to a single conventional superconductor, but repulsive interactions are also necessary to reach the topological phase in these setups, which require a strength of induced crossed Andreev (interwire)  pairing exceeding that of the direct (intrawire) pairing \cite{Reeg:2017,Recher:2002,Bena:2002}. While it would be beneficial to engineer a DIII topological superconductor in a noninteracting 1D system coupled to a single conventional superconductor, as such a setup could avoid relying on unconventional superconductivity or interactions that are difficult to control experimentally, it was recently shown that this is not possible in a fully time-reversal invariant system \cite{Haim:2016}.

In this paper, we show that such a 1D topological superconductor in class DIII can be realized when time-reversal symmetry is explicitly broken. While the full Hamiltonian (describing the 1D system, the superconductor, and the tunnel coupling) possesses only particle-hole symmetry and is thus in symmetry class D, it is possible to place the system in symmetry class DIII after integrating out the superconductor \cite{Sau:2010prox,Potter:2011,Kopnin:2011,Zyuzin:2013,vanHeck:2016,Reeg:2017_2,Reeg:2017,Reeg:2017_3} and projecting to an effective 1D model \cite{note}. We establish three necessary criteria to realize a DIII topological phase. First, the 1D system must obey an ``emergent" time-reversal symmetry. That is, given the Hamiltonian density $h_k$ of the 1D system, there must exist a unitary matrix $\mathcal{T}_\od$ such that $\mathcal{T}_\od^\dagger h_k\mathcal{T}_\od=h_{-k}^*$ and $\mathcal{T}_\od^2=-1$. [While a specific example could be the physical time-reversal symmetry $\mathcal{T}_\od=i\sigma_y$, where $\sigma_{x,y,z}$ is a Pauli matrix acting in spin space, we do not restrict ourselves to this case.] Second, the self-energy induced on the 1D system by the superconductor must preserve the emergent time-reversal symmetry. Third, the anomalous (pairing) component of the self-energy must have both positive and negative eigenvalues.

After a general discussion, we provide a simple  model which realizes the DIII topological phase in 1D. We consider a system of two Rashba nanowires with opposite Zeeman splittings coupled to an $s$-wave superconductor. We show that such a system undergoes a topological phase transition under certain conditions. By explicitly solving for the wave functions of the Majorana bound states, we show that the topological phase is characterized by the presence of a Kramers pair of Majorana fermions that is protected by the emergent time-reversal symmetry.

\paragraph*{Minimum requirements for DIII topological phase.} We consider a general 1D (noninteracting) system coupled to a conventional superconductor. We assume that the system is translationally invariant along the direction of the 1D system, allowing us to define a conserved momentum $k$. The Hamiltonian of the 1D system is given by
\begin{equation} \label{H1D}
H_\od=\int\frac{dk}{2\pi}\psi_k^\dagger h_k\psi_k,
\end{equation}
where $\psi_k$ is a spinor annihilation operator acting on all internal degrees of freedom of the 1D system (spin, subband, etc.) and $h_k$ is a Hermitian matrix. We assume that there exists a unitary matrix $\mathcal{T}_\od$ which acts as an effective time-reversal symmetry on the Hamiltonian, such that $\mathcal{T}_\od^\dagger h_k\mathcal{T}_\od=h^*_{-k}$ and $\mathcal{T}_\od^2=-1$. Introducing the Nambu spinor $\Psi_k^\dagger=(\psi_k^\dagger,\psi_{-k}^T\mathcal{T}_\od)$, the Hamiltonian Eq.~\eqref{H1D} can be expressed as
\begin{equation} \label{H1DNambu}
H_\od=\frac{1}{2}\int\frac{dk}{2\pi}\,\Psi_k^\dagger\mathcal{H}_k^\od\Psi_k,
\end{equation}
where $\mathcal{H}_k=\tau_zh_k$ and $\tau_{x,y,z}$ is a Pauli matrix acting in Nambu space.

The 1D system is coupled to a conventional superconductor which is described by a BCS Hamiltonian,
\begin{equation} \label{Hsc}
H_\text{sc}=\frac{1}{2}\int\frac{dk}{2\pi}\int d\x_\perp\,\eta^\dagger_k(\xp)\mathcal{H}_k^\text{sc}(\xp)\eta_k(\xp),
\end{equation}
where $\xp$ denotes directions transverse to the 1D system and $\eta_k^\dagger=(\eta_{k\uparrow}^\dagger,\eta_{k\downarrow}^\dagger,-\eta_{-k\downarrow},\eta_{-k\uparrow})$ is a spinor creation operator in Nambu $\otimes$ spin space. The Hamiltonian density in this basis is given by
$\mathcal{H}_k^\text{sc}(\xp)=\tau_z[(k^2-\nabla_\perp^2)/2m_\text{sc}-\mu_\text{sc}]+\tau_x\Delta$, where $m_\text{sc}$, $\mu_\text{sc}$, and $\Delta$ are the effective mass, chemical potential, and pairing potential of the superconductor, respectively. The superconductor is time-reversal invariant, $\mathcal{T}_\text{sc}^\dagger\mathcal{H}_k^\text{sc}\mathcal{T}_\text{sc}=\mathcal{H}_{-k}^{\text{sc}*}$, where $\mathcal{T}_\text{sc}=i\sigma_y$ is the physical time-reversal operator, with $\sigma_{x,y,z}$ the Pauli
matrix acting in spin space.

We allow for a linear coupling between the 1D system and the superconductor of the form
\begin{equation} \label{Ht}
H_t=\frac{1}{2}\int\frac{dk}{2\pi}\int d\xp[\eta_k^\dagger(\xp)T_k(\xp)\Psi_k+H.c.].
\end{equation}
The tunneling matrix in Nambu space, $T_k(\xp)$, can be expressed generally as
\begin{equation} \label{tunneling}
T_k=T_k^0+\tau_zT_k^z,
\end{equation}
where $T_k^0=[t_k-\mathcal{T}_\text{sc}^\dagger t_{-k}^*\mathcal{T}_\od]/2$, $T_k^z=[t_k+\mathcal{T}_\text{sc}^\dagger t_{-k}^*\mathcal{T}_\od]/2$, and $t_k$ is a tunneling matrix acting on all additional degrees of freedom in the system \cite{note4}. Combining Eqs.~\eqref{H1DNambu}-\eqref{Ht}, the full Hamiltonian can be expressed as
\begin{equation} \label{fullH}
H=\frac{1}{2}\int\frac{dk}{2\pi}\begin{pmatrix} \Psi_k^\dagger & \eta_k^\dagger \end{pmatrix}
	\begin{pmatrix} \mathcal{H}_k^\od & T_k^\dagger \\ T_k & \mathcal{H}_k^\text{sc} \end{pmatrix}
	\begin{pmatrix} \Psi_k \\ \eta_k \end{pmatrix},
\end{equation}
where we suppress explicit reference to $\xp$ for brevity. Note that the time-reversal symmetry of the full Hamiltonian is broken by having $T_k^0\neq0$; i.e., $\mathcal{T}^\dagger\mathcal{H}_k\mathcal{T}\neq\mathcal{H}_{-k}^*$, where $\mathcal{H}_k$ is the Hamiltonian density of Eq.~\eqref{fullH} and $\mathcal{T}=\text{diag}(\mathcal{T}_\od,\mathcal{T}_\text{sc})$. For this reason, it was assumed that $T_k^0=0$ in Ref.~\cite{Haim:2016}.

We now project our system to an effective 1D model by integrating out the superconductor \cite{Sau:2010prox,Potter:2011,Kopnin:2011,Zyuzin:2013,vanHeck:2016,Reeg:2017_2,Reeg:2017,Reeg:2017_3}. The superconductor induces a self-energy on the 1D system given by
\begin{equation} \label{selfenergy}
\Sigma_k(\omega)=\int d\xp\int d\xp'\,T_k^\dagger(\xp) G_{k,\omega}^\text{sc}(\xp,\xp')T_k(\xp'),
\end{equation}
where $G_{k,\omega}^\text{sc}(\xp,\xp')$ is the Matsubara Green's function of the bare superconductor, defined such that $[i\omega-\mathcal{H}_k^\text{sc}(\xp)]G_{k,\omega}^\text{sc}(\xp,\xp')=\delta(\xp-\xp')$. In the limit of weak tunneling, where the relevant pairing energies in the 1D system are $\omega\ll\Delta$, it is sufficient to evaluate the self-energy at $\omega=0$. In this case, the system is described by an effective 1D Hamiltonian given by $\mathcal{H}_k^\text{eff}=\mathcal{H}_k^\od+\Sigma_k$. Since it was already assumed that $\mathcal{H}_k^\od$ obeys an effective time-reversal symmetry, the Hamiltonian $\mathcal{H}_k^\text{eff}$ is in class DIII if the self-energy of Eq.~\eqref{selfenergy} preserves this symmetry,
\begin{equation} \label{selfenergycondition}
\mathcal{T}_\od^\dagger\Sigma_k\mathcal{T}_\od=\Sigma_{-k}^*.
\end{equation}
Hence, $\mathcal{T}_\od$ acts as an emergent time-reversal symmetry which exists only in the low-dimensional subspace. Assuming that the self-energy satisfies Eq.~\eqref{selfenergycondition}, we can decompose it into normal and anomalous parts as 
\begin{equation} \label{selfenergydecomposition}
\Sigma_k=\tau_z\Sigma_k^N+\tau_x\Sigma_k^A,
\end{equation}
where $\Sigma_k^N=G_k^N(T_k^{0\dagger}T_k^0+T_k^{z\dagger}T_k^z)$ and $\Sigma_k^A=G_k^A(T_k^{0\dagger}T_k^0-T_k^{z\dagger}T_k^z)$ \cite{supp}. In arriving at Eq.~\eqref{selfenergydecomposition}, we have utilized the fact that the superconducting Green's function can be similarly decomposed as $G_k^\text{sc}=\tau_zG_k^N+\tau_xG_k^A$, where $G_k^N$ and $G_k^A$ are scalars.

The anomalous self-energy, which represents the induced pairing in the 1D system, can be expressed in a form
\begin{equation} \label{anomalous}
\Sigma_k^A=\Sigma_k^z-\Sigma_k^0,
\end{equation}
where $\Sigma_k^i=-G_k^AT_k^{i\dagger}T_k^i$. It was shown in Ref.~\cite{Haim:2016} that the class DIII topological invariant can only take a nontrivial value if $\Sigma_k^A$ has both positive and negative eigenvalues. It was also shown that $\Sigma_k^z$ is always positive semidefinite; hence, if the tunneling Hamiltonian is time-reversal invariant with $\Sigma_k^0=0$, it is not possible to realize a topological phase. By extension, it is straightforward to show that $\Sigma_k^0$ must also be positive semidefinite. Consequently, the topological invariant must also take a trivial value if $\Sigma_k^z=0$ (in which case $\Sigma_k^A$ is negative semidefinite). However, if both terms in Eq.~\eqref{anomalous} are nonzero, $\Sigma_k^A$ is not restricted to be either positive semidefinite or negative semidefinite, and it is possible to have a topologically nontrivial phase.

We have thus established three minimal criteria to realize a class DIII topological phase in a noninteracting 1D system: the bare 1D system must obey an effective time-reversal symmetry, the self-energy induced by the superconductor must preserve this symmetry, and the anomalous self-energy must have both positive and negative eigenvalues. The final requirement can be satisfied only if the full tunneling Hamiltonian [Eq.~\eqref{fullH}] is not time-reversal invariant. We will now provide a  model which satisfies all three criteria, showing indeed that the class DIII topological phase can be realized.

\begin{figure}[t!]
\centering
\includegraphics[width=0.8\linewidth]{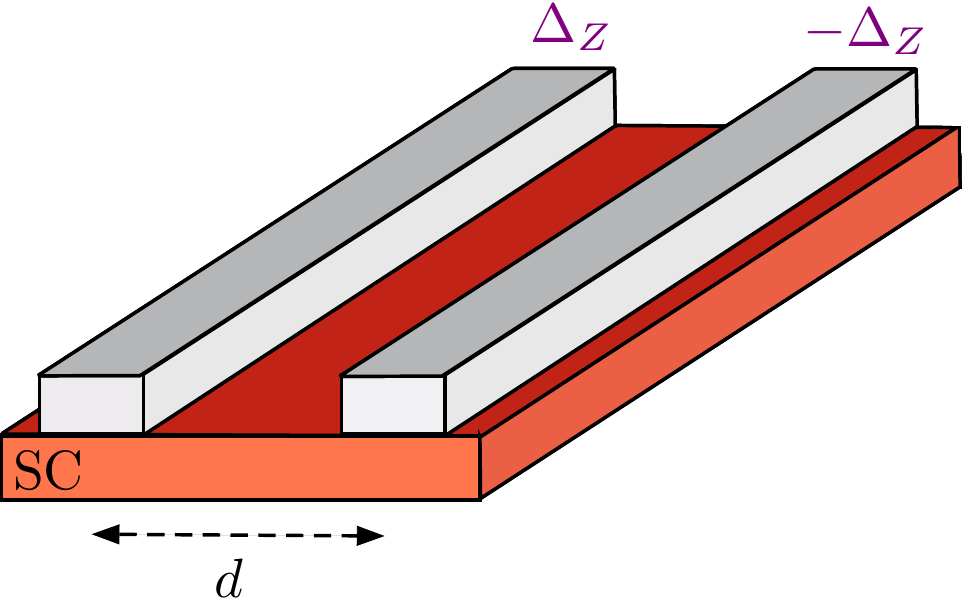}
\caption{\label{setup} Two nanowires, separated by a distance $d$, are coupled to an infinite superconducting plane. The two nanowires are assumed to have opposite Zeeman splittings.}
\end{figure}

\paragraph*{Model.}
We consider the geometry shown in Fig.~\ref{setup}. Two Rashba nanowires, separated by a distance $d$ (let us take one wire to be located at $z=0$ and the other at $z=d$), are coupled to an infinite 2D $s$-wave superconducting plane. We take the two nanowires to have opposite Zeeman splitting, which can be achieved by applying an antiparallel external magnetic field to each wire or by applying a uniform external magnetic field to two wires with opposite $g$-factors.

The two nanowires are described by the Hamiltonian density
\begin{equation} \label{modelh}
h_k=\xi_k-\alpha k\sigma_z-\Delta_Z\eta_z\sigma_x,
\end{equation}
where $\xi_k=k^2/2m_w-\mu_w$ ($m_w$ is the effective mass and $\mu_w$ the chemical potential of the nanowires), $\alpha$ is the Rashba spin-orbit interaction constant (we choose our spin quantization axis along the direction of the effective Rashba field), and $\Delta_Z=g\mu_BB/2$ is the Zeeman splitting in an external magnetic field of strength $B$ ($g$ is the nanowire $g$-factor and $\mu_B$ is the Bohr magneton). The Pauli matrix $\eta_{x,y,z}$ acts in left/right wire space. Crucially, we impose that the two nanowires are identical, with only a change in the sign of the Zeeman splitting. Although the Zeeman term in Eq.~\eqref{modelh} explicitly breaks time-reversal, the Hamiltonian density obeys an effective time-reversal symmetry $\mathcal{T}_\od^\dagger h_k\mathcal{T}_\od=h_{-k}^*$, where $\mathcal{T}_\od=i\eta_x\sigma_y$ and $\mathcal{T}_\od^2=-1$.

The self-energy induced on the two nanowires by the superconductor is given in Eq.~\eqref{selfenergy}. Assuming that $\mu_s\gg\mu_w$, we evaluate the Green's function of the bulk 2D superconductor for momenta $k\ll k_{Fs}$ ($k_{Fs}=\sqrt{2m_s\mu_s}$ is the Fermi momentum of the superconductor) to give
\begin{equation}
\begin{aligned}
G^\text{sc}_{0,0}(z,z')&=-\frac{1}{v_{Fs}}[\tau_x\cos(k_{Fs}|z-z'|) \\
	&-\tau_z\sin(k_{Fs}|z-z'|)]e^{-|z-z'|/\xi_s},
\end{aligned}
\end{equation}
where $\xi_s=v_{Fs}/\Delta$ is the coherence length and $v_{Fs}=k_{Fs}/m_s$ the Fermi velocity of the superconductor (we have also expanded in the limit $\mu_s\gg\Delta$). We assume local spin- and momentum-independent tunneling of the form $t_k(z)=\begin{bmatrix} t\delta(z) & t\delta(z-d)\end{bmatrix}$, where $t$ is a (scalar) tunneling amplitude which has the same strength in both nanowires. This gives $T_k^{0}=t_k(1-\eta_x)/2$ and $T_k^z=t_k(1+\eta_x)/2$. Evaluating the self-energy Eq.~\eqref{selfenergy}, we find
\begin{equation}
\Sigma=\Gamma\tau_z\eta_x+\tau_x(\Delta_c+\Delta_d\eta_x),
\end{equation}
where we define a single-particle interwire tunnel coupling $\Gamma=\gamma\sin(k_{Fs}d)e^{-d/\xi_s}$, an induced direct (intrawire) pairing potential $\Delta_d=\gamma$, and an induced crossed Andreev (interwire) pairing potential $\Delta_c=\gamma\cos(k_{Fs}d)e^{-d/\xi_s}$ \cite{Reeg:2017}. Here, $\gamma=t^2/v_{Fs}$ is a tunneling energy scale. Note that the pairing potentials always satisfy $\Delta_d>\Delta_c$.

Taking into account the self-energy, the effective Hamiltonian describing the double nanowire system is given by
\begin{equation} \label{Heff2}
\mathcal{H}^\text{eff}_k=\tau_z(\xi_k-\alpha k\sigma_z-\Delta_Z\eta_z\sigma_x+\Gamma\eta_x)+\tau_x(\Delta_c+\Delta_d\eta_x).
\end{equation}
Because the self-energy preserves the effective time-reversal symmetry, $\mathcal{T}_\od^\dagger\Sigma\mathcal{T}_\od=\Sigma^*$, we also have $\mathcal{T}_\od^\dagger\mathcal{H}_k^\text{eff}\mathcal{T}_\od=\mathcal{H}_{-k}^{\text{eff}*}$. Additionally, the effective Hamiltonian possesses a particle-hole symmetry $\mathcal{P}^\dagger\mathcal{H}_k^\text{eff}\mathcal{P}=-\mathcal{H}_{-k}^{\text{eff}*}$, where $\mathcal{P}=\tau_y\eta_x\sigma_y$ is a unitary matrix satisfying $\mathcal{P}^2=1$. Finally, the effective Hamiltonian possesses a chiral symmetry $\{\mathcal{C},\mathcal{H}_k^\text{eff}\}=0$, where $\mathcal{C}=\mathcal{T}_{\od}\mathcal{P}=i\tau_y$. These three properties of the Hamiltonian place it in the DIII symmetry class. Additionally, the anomalous self-energy $\Sigma^A=\Delta_c+\Delta_d\eta_x$ has both positive and negative eigenvalues, $\Delta_c\pm\Delta_d$. Because all of our previously established criteria are met in this setup, it is possible to have a topological phase.

\begin{figure}[t!]
\centering
\includegraphics[width=0.9\linewidth]{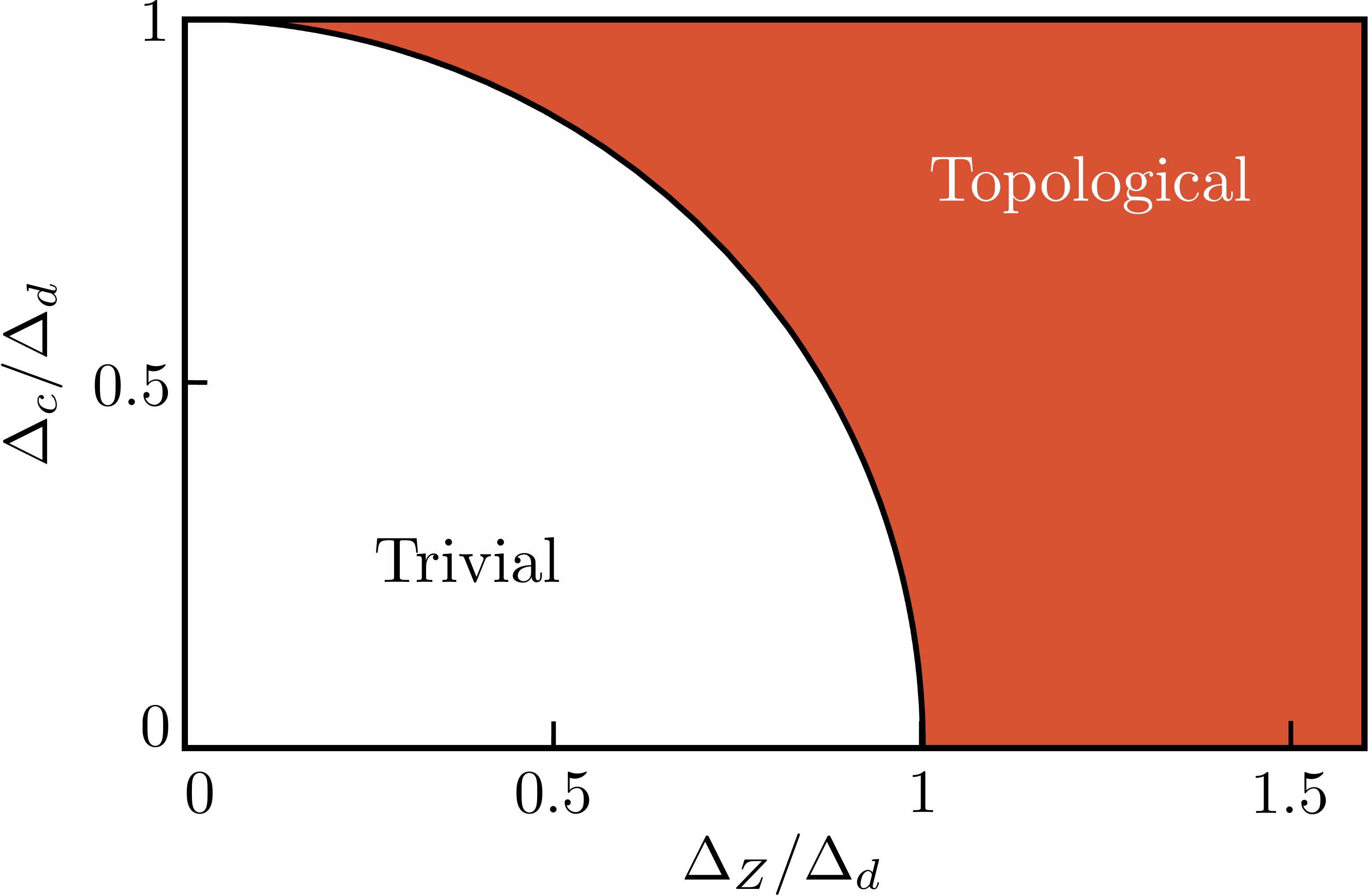}
\caption{\label{phasediagram} Phase diagram of our  model. A phase boundary at $\Delta_Z^2=\Delta_d^2-\Delta_c^2$ separates trivial and DIII topological superconducting phases. The topological phase is characterized by the presence of a Kramers pair of Majorana fermions that is protected by the emergent time-reversal symmetry.}
\end{figure}

To determine whether such a topological phase exists in this setup, we search for a $k=0$ gap-closing transition by enforcing $\det(\mathcal{H}_0^{\text{eff}})=0$. We find
\begin{equation}
\begin{aligned}
&\det(\mathcal{H}_0^\text{eff})=\bigl\{\Delta_Z^4+2\Delta_Z^2(\Gamma^2-\Delta_d^2+\Delta_c^2-\mu_w^2) \\
	&+[(\Delta_d-\Delta_c)^2+(\Gamma-\mu_w)^2][(\Delta_d+\Delta_c)^2+(\Gamma+\mu_w)^2]\bigr\}^2,
\end{aligned}
\end{equation}
which yields a gap-closing at the critical Zeeman splitting $(\Delta_Z^c)^2=\Delta_d^2-\Delta_c^2+\mu_w^2-\Gamma^2\pm2i(\mu_w\Delta_c-\Gamma\Delta_d)$. Therefore, in order to have a physical transition, the chemical potential of the nanowires must be tuned to $\mu_w=\Gamma\Delta_d/\Delta_c$. For simplicity, let us assume that the system is tuned in such a way that the interwire tunnel coupling vanishes, $\Gamma=0$ [which can be done by tuning $\sin(k_{Fs}d)=0$]. In this case, the critical Zeeman splitting at which the gap closes is given by
\begin{equation}
\Delta_Z^c=\sqrt{\Delta_d^2-\Delta_c^2}.
\end{equation}
The phase diagram of our model is displayed in Fig.~\ref{phasediagram}. We find two distinct phases whose topological characterization can be inferred along the line $\Delta_c=0$, corresponding to the case when the two wires are decoupled ($d\gg\xi_s$). For $\Delta_Z<\Delta_d$, both wires are in a topologically trivial phase. For $\Delta_Z>\Delta_d$, both wires are in a topologically nontrivial phase, with each wire hosting its own distinct pair of Majorana bound states. Because the number of Majorana bound states is a topological invariant that cannot be changed without closing the gap, we conclude that $\Delta_Z^2<\Delta_d^2-\Delta_c^2$ corresponds to a topologically trivial phase while $\Delta_Z^2>\Delta_d^2-\Delta_c^2$ corresponds to a topologically nontrivial phase with two pairs of Majorana bound states.

To further establish the presence of a topologically nontrivial phase, we explicitly solve for the wave functions of the Majorana bound states. We now take our nanowires to be semi-infinite ($x>0$), and we assume that the effective Hamiltonian of Eq.~\eqref{Heff2} remains valid for the semi-infinite case after replacing $k\to-i\partial_x$ \cite{note3}. States in the nanowires obey a Bogoliubov-de Gennes equation given by $\mathcal{H}_\text{eff}(x)\phi(x)=E\phi(x)$. By constructing a general zero-energy solution and imposing a vanishing boundary condition $\phi(0)=0$, we find two Majorana bound state solutions in the topological phase ($\Delta_Z^2>\Delta_d^2-\Delta_c^2$) and no solutions in the trivial phase ($\Delta_Z^2<\Delta_d^2-\Delta_c^2$). The full analytical expressions for the wave functions are given in the limit of strong spin-orbit interaction ($m_w\alpha^2\gg\Delta_d,\Delta_c,\Delta_Z$) in the Supplemental Material \cite{supp}. We find that the two Majorana wave functions in the topological phase are orthogonal,
\begin{equation}
\phi_2^\dagger(x)\phi_1(x)=0,
\end{equation}
and related by the effective time-reversal symmetry,
\begin{equation}
\begin{aligned}
\phi_1(x)&=\mathcal{T}_\od\phi_2^*(x), \\
\phi_2(x)&=-\mathcal{T}_\od\phi_1^*(x).
\end{aligned}
\end{equation}
Hence, the two Majorana bound states in the system form a Kramers pair that is protected by the emergent time-reversal symmetry.

\paragraph*{Conclusions.}
We have shown that a topological superconductor in symmetry class DIII can be realized in a noninteracting 1D system proximity coupled to a conventional superconductor. Crucially, the full Hamiltonian (incorporating the 1D system, the parent superconductor, and the tunneling term) must not possess an effective time-reversal symmetry, with such a symmetry emerging only after projection to an effective 1D model. We provide an explicit example realizing such a class DIII topological superconductor, showing that the topological phase is characterized by a Kramers pair of Majorana bound states which is protected by the effective time-reversal symmetry of the system. We believe that our general criteria can be applied to realize class DIII topological superconductivity in a multitude of additional systems coupled to a bulk $s$-wave superconductor, for example in nanowires with helical magnetization of opposite helicity, antiferromagnetically coupled spin chains, magnetic topological insulators with opposite magnetization, or ferromagnetic atomic chains with opposite magnetization.

\paragraph*{Acknowledgments.} This work was supported by the Swiss National Science Foundation and the NCCR QSIT.

\paragraph*{Note.} Upon completion of this manuscript, we became aware of a very recent preprint, Ref.~\cite{Huang:2017}, which discusses the realization of a class DIII topological phase in 2D antiferromagnetic quantum spin Hall insulators.

\bibliography{bibDIII}

\begin{thebibliography}{54}
\bibitem [{not({\natexlab{b}})}]{note4}%
  \BibitemOpen
  \href@noop {} {}  \bibinfo {note} {Note that we do not
  consider the possibility that $t_k$ is a two-body operator as in
  Ref.~\cite{Schrade:2015}, as this corresponds to an interacting
  model.}\BibitemShut {Stop}%
\bibitem [{not({\natexlab{c}})}]{note3}%
  \BibitemOpen
  \href@noop {} {}  \bibinfo {note} {Because the proximity
  effect was treated in the bulk of the nanowires, there could be additional
  boundary effects that are not properly accounted for in the effective
  Hamiltonian. However, the presence or absence of topological Majorana modes
  is insensitive to such boundary effects.}\BibitemShut {Stop}%
\end{thebibliography}

\newpage

\begin{widetext}

\setcounter{equation}{0}
\setcounter{figure}{0}
\renewcommand{\theequation}{S\arabic{equation}}
\renewcommand{\thefigure}{S\arabic{figure}}

\section{Supplemental Material}
\subsection{Form of Self-Energy}
In this section, we prove that the self-energy can be decomposed into normal and anomalous components as in Eq.~\eqref{selfenergydecomposition} of the main text. The self-energy of the effective 1D model is given by
\begin{equation}
\Sigma_k(\omega)=\int d\xp\int d\xp'\,T_k^\dagger(\xp) G_{k,\omega}^\text{sc}(\xp,\xp')T_k(\xp').
\end{equation}
Substituting the tunneling term $T_k=T_k^0+T_k^z$, and noting that the superconducting Green's function (at $\omega=0$) can be decomposed into normal and anomalous parts as $G_k^\text{sc}=\tau_zG_k^N+\tau_xG_k^A$, where $G_k^{N}$ and $G_k^A$ are scalars, we find a self-energy given by (we suppress explicit reference to the transverse coordinates $\xp$ and $\xp'$ for brevity)
\begin{equation} \label{selfenergyexpansion}
\begin{aligned}
\Sigma_k&=G_k^N[\tau_z(T_k^{0\dagger}T_k^0+T_k^{z\dagger}T_k^z)+\tau_0(T_k^{0\dagger}T_k^z+T_k^{z\dagger}T_k^0)] +G_k^A[\tau_x(T_k^{0\dagger}T_k^0-T_k^{z\dagger}T_k^z)-i\tau_y(T_k^{0\dagger}T_k^z-T_k^{z\dagger}T_k^0)].
\end{aligned}
\end{equation}
We now require the self-energy to preserve the effective time-reversal symmetry of the 1D system by enforcing $\mathcal{T}_\od^\dagger\Sigma_k\mathcal{T}_\od=\Sigma_{-k}^*$. Because the four terms of Eq.~\eqref{selfenergyexpansion} are linearly independent, each term must separately obey this condition. 

Let us first consider the term proportional to $\tau_0$; expanding out this term using the definitions $T_k^{0}=[t_k-\mathcal{T}_\text{sc}^\dagger t_{-k}^*\mathcal{T}_\od]/2$ and $T_k^z=[t_k+\mathcal{T}_\text{sc}^\dagger t_{-k}^*\mathcal{T}_\od]/2$ gives
\begin{equation} \label{tau0}
T_k^{0\dagger}T_k^z+T_k^{z\dagger}T_k^0=\frac{1}{2}(t_k^\dagger t_k-\mathcal{T}_\od^\dagger t_{-k}^Tt_{-k}^*\mathcal{T}_\od).
\end{equation}
Applying the time-reversal operator to Eq.~\eqref{tau0} and using the fact that $\mathcal{T}_\od^2=-1$, we see that in order to preserve the time-reversal symmetry of the 1D system, the self-energy must satisfy
\begin{equation}
\mathcal{T}_\od^\dagger t_k^\dagger t_k\mathcal{T}_\od-t_{-k}^Tt_{-k}^*=t_{-k}^Tt_{-k}^*-\mathcal{T}_\od^\dagger t_k^\dagger t_k\mathcal{T}_\od,
\end{equation}
or, equivalently,
\begin{equation} \label{relation1}
t_k^\dagger t_k=\mathcal{T}_\od^\dagger t_{-k}^T t_{-k}^*\mathcal{T}_\od.
\end{equation}
Hence, the term proportional to $\tau_0$ in Eq.~\eqref{selfenergyexpansion} must vanish.

Next, we consider the term proportional to $\tau_y$ in Eq.~\eqref{selfenergyexpansion},
\begin{equation} \label{tauy}
T_k^{0\dagger}T_k^z-T_k^{z\dagger}T_k^0=\frac{1}{2}(t_k^\dagger\mathcal{T}_\text{sc}^\dagger t_{-k}^*\mathcal{T}_\od-\mathcal{T}_\od^\dagger t_{-k}^T\mathcal{T}_\text{sc}t_k).
\end{equation}
Applying the time-reversal operator to Eq.~\eqref{tauy}, we require
\begin{equation}
t_{-k}^T\mathcal{T}_\text{sc}t_k\mathcal{T}_\od-\mathcal{T}_\od^\dagger t_k^\dagger\mathcal{T}_\text{sc}^\dagger t_{-k}^*=t_{-k}^T\mathcal{T}_\text{sc}^\dagger t_k\mathcal{T}_\od-\mathcal{T}_\od^\dagger t_k^\dagger\mathcal{T}_\text{sc} t_{-k}^*.
\end{equation}
However, because $\mathcal{T}_\text{sc}^\dagger=-\mathcal{T}_\text{sc}$, this amounts to
\begin{equation} \label{relation2}
t_k^\dagger\mathcal{T}_\text{sc}^\dagger t_{-k}^*\mathcal{T}_\od=\mathcal{T}_\od^\dagger t_{-k}^T\mathcal{T}_\text{sc}t_k,
\end{equation}
meaning that the product $t_k^\dagger\mathcal{T}_\text{sc}^\dagger t_{-k}^*\mathcal{T}_\od$ is necessarily Hermitian and that the term proportional to $\tau_y$ in Eq.~\eqref{selfenergyexpansion} must also vanish.

Using the relations defined in Eqs.~\eqref{relation1} and \eqref{relation2}, we see that the terms proportional to $\tau_z$ and $\tau_x$ in Eq.~\eqref{selfenergyexpansion} do not vanish in general. Hence, the self-energy can be expressed in the form
\begin{equation}
\Sigma_k=G_k^N\tau_z(T_k^{0\dagger}T_k^0+T_k^{z\dagger}T_k^z)+G_k^A\tau_x(T_k^{0\dagger}T_k^0-T_k^{z\dagger}T_k^z),
\end{equation}
as given in Eq.~\eqref{selfenergydecomposition} of the main text.

\subsection{Majorana Wave Functions}
In this section, we explicitly solve for the Majorana wave functions in a semi-infinite geometry. States in the nanowire obey a Bogoliubov-de Gennes equation given by
\begin{equation} \label{BdG}
\mathcal{H}_\text{eff}(x)\phi(x)=E\phi(x),
\end{equation}
where $\phi(x)$ is a spinor wave function and $\mathcal{H}_\text{eff}(x)$ is given by
\begin{equation}
\mathcal{H}_\text{eff}(x)=\tau_z(-\partial_x^2/2m_w+i\alpha \partial_x\sigma_z-\Delta_Z\eta_z\sigma_x)+\tau_x(\Delta_c+\Delta_d\eta_x).
\end{equation}
Because the wires are semi-infinite, Majorana solutions are necessarily at zero energy. Setting $E=0$, we rewrite Eq.~\eqref{BdG} in the form
\begin{equation}
\partial_x\tilde\phi(x)=M\tilde\phi(x),
\end{equation}
where $\tilde\phi=(\partial_x\phi,\phi)^T$ and
\begin{equation}
M=\begin{pmatrix}
	2m_wi\alpha\sigma_z & -2m_w[\Delta_Z\eta_z\sigma_x+i\tau_y(\Delta_c+\Delta_d\eta_x)] \\
	\mathbb{I}_{8\times8} & 0 
	\end{pmatrix}.
\end{equation}
Any zero-energy solution to Eq.~\eqref{BdG} can be written in the form
\begin{equation} \label{WFexpansion}
\phi(x)=\sum_nc_n\chi_ne^{ik_nx},
\end{equation}
where $ik_n$ are the eigenvalues of $M$ and $\chi_n$ are the corresponding eigenvectors. 

Eigenvalues of $M$ must satisfy (for brevity, we temporarily set $2m_w=1$)
\begin{equation}
\begin{aligned}
0&=[k^8-2\alpha^2k^6+k^4(\alpha^4+2\Delta_d^2+2\Delta_c^2-2\Delta_Z^2)+8\alpha k^3\Delta_d\Delta_c+2\alpha^2k^2(\Delta_d^2+\Delta_c^2+\Delta_Z^2)+(\Delta_Z^2-\Delta_d^2+\Delta_c^2)^2] \\
	&\times[k^8-2\alpha^2k^6+k^4(\alpha^4+2\Delta_d^2+2\Delta_c^2-2\Delta_Z^2)-8\alpha k^3\Delta_d\Delta_c+2\alpha^2k^2(\Delta_d^2+\Delta_c^2+\Delta_Z^2)+(\Delta_Z^2-\Delta_d^2+\Delta_c^2)^2].
\end{aligned}
\end{equation}
While this equation cannot be solved analytically in general, we can solve it in the limit of strong spin-orbit interaction, $\alpha^2\gg\Delta_Z,\Delta_d,\Delta_c$. First, we expand in the vicinity of the exterior branches of the spectrum by setting $k=\pm\alpha+\delta k_e$ and expanding for $\delta k_e\ll\alpha$. We find
\begin{equation}
[(\Delta_d+\Delta_c)^2+\alpha^2\delta k_e^2][(\Delta_d-\Delta_c)^2+\alpha^2\delta k_e^2]=0.
\end{equation}
Therefore, at both the positive and negative exterior branches, we find four allowed momenta with $\delta k_e=\pm i(\Delta_d\pm\Delta_c)/\alpha$. Next, we expand in the vicinity of the interior branches of the spectrum by setting $k=\delta k_i$ and again expanding for $\delta k_i\ll\alpha$; we find
\begin{equation}
[\alpha^4\delta k_i^4+2\alpha^2(\Delta_Z^2+\Delta_d^2+\Delta_c^2)\delta k_i^2+(\Delta_Z^2-\Delta_d^2+\Delta_c^2)^2]^2=0.
\end{equation}
For the interior branches, we find four doubly degenerate momenta corresponding to $\delta k_i=\pm\frac{i}{\alpha}(\sqrt{\Delta_Z^2+\Delta_c^2}\pm\Delta_d)$. 

For bound states, only momenta with positive imaginary part are allowed. Restoring the factors of $2m_w$, the allowed momenta in the topological phase ($\Delta_Z^2>\Delta_d^2-\Delta_c^2$) are given by
\begin{equation} \label{momenta}
\begin{aligned}
k_1&=2m_w\alpha+i(\Delta_d+\Delta_c)/\alpha, \\
k_2&=2m_w\alpha+i(\Delta_d-\Delta_c)/\alpha, \\
k_3&=-2m_w\alpha+i(\Delta_d+\Delta_c)/\alpha, \\
k_4&=-2m_w\alpha+i(\Delta_d-\Delta_c)/\alpha, \\
k_5=k_6&=i\left(\sqrt{\Delta_Z^2+\Delta_c^2}+\Delta_d\right)/\alpha, \\
k_7=k_8&=i\left(\sqrt{\Delta_Z^2+\Delta_c^2}-\Delta_d\right)/\alpha.
\end{aligned}
\end{equation}
Substituting Eqs.~\eqref{momenta} and \eqref{WFexpansion} into Eq.~\eqref{BdG} and similarly expanding in the limit of strong spin-orbit interaction, we find the corresponding eigenvectors
\begin{align}
\chi_1&=\frac{1}{2}(-1,0,-1,0,-i,0,-i,0)^T, \nonumber \\
\chi_2&=\frac{1}{2}(-i,0,i,0,-1,0,1,0)^T, \nonumber \\
\chi_3&=\frac{1}{2}(0,i,0,i,0,1,0,1)^T, \nonumber \\
\chi_4&=\frac{1}{2}(0,-i,0,i,0,1,0,-1)^T, \\
\chi_5&=\frac{1}{2\sqrt{\Delta_Z^2+\beta^2}}(\beta,-i\Delta_Z,\beta,i\Delta_Z,-i\beta,-\Delta_Z,-i\beta,\Delta_Z)^T, \nonumber \\
\chi_6&=\frac{1}{2\sqrt{\Delta_Z^2+\beta^2}}(\Delta_Z,-i\beta,-\Delta_Z,-i\beta,i\Delta_Z,\beta,-i\Delta_Z,\beta)^T, \nonumber \\
\chi_7&=\frac{-1}{2\sqrt{\Delta_Z^2+\beta^2}}(-\beta,i\Delta_Z,\beta,i\Delta_Z,i\beta,\Delta_Z,-i\beta,\Delta_Z)^T, \nonumber \\
\chi_8&=\frac{1}{2\sqrt{\Delta_Z^2+\beta^2}}(i\Delta_Z,\beta,i\Delta_Z,-\beta,-\Delta_Z,i\beta,-\Delta_Z,-i\beta)^T, \nonumber
\end{align}
where we define $\beta=\sqrt{\Delta_Z^2+\Delta_c^2}+\Delta_c$.

The coefficients $c_{1-8}$ are determined by imposing the boundary condition $\phi(0)=0$. We find two solutions given by
\begin{equation} \label{WFs}
\begin{aligned}
\phi_1(x)&=\mathcal{N}_1\biggl(\frac{i\Delta_Z}{\sqrt{\Delta_Z^2+\beta^2}}\chi_1e^{ik_1x}-\frac{i\beta}{\sqrt{\Delta_Z^2+\beta^2}}\chi_4e^{ik_4x}+\chi_8e^{ik_8x}\biggr), \\
\phi_2(x)&=\mathcal{N}_2\biggl(-\frac{i\beta}{\sqrt{\Delta_Z^2+\Delta_c^2}}\chi_2e^{ik_2x}+\frac{\Delta_Z}{\sqrt{\Delta_Z^2+\Delta_c^2}}\chi_3e^{ik_3x}+\chi_7e^{ik_7x}\biggr),
\end{aligned}
\end{equation}
where $\mathcal{N}_{1(2)}$ are normalization constants. The wave functions are normalized according to
\begin{equation}
\int_0^\infty dx\,\phi^\dagger_i(x)\phi_i(x)=2.
\end{equation}
Evaluating the integral, we find normalization constants
\begin{equation} \label{normalization}
\begin{aligned}
\mathcal{N}_1&=\frac{1}{\sqrt{\alpha}}\biggl(\frac{1}{\sqrt{\Delta_Z^2+\Delta_c^2}-\Delta_d}+\frac{\beta^2}{(\Delta_d-\Delta_c)(\Delta_Z^2+\beta^2)}+\frac{\Delta_Z^2}{(\Delta_d+\Delta_c)(\Delta_Z^2+\beta^2)}\biggr)^{-1/2}, \\
\mathcal{N}_2&=\frac{1}{\sqrt{\alpha}}\biggl(\frac{1}{\sqrt{\Delta_Z^2+\Delta_c^2}-\Delta_d}+\frac{\beta^2}{(\Delta_d-\Delta_c)(\Delta_Z^2+\Delta_c^2)}+\frac{\Delta_Z^2}{(\Delta_d+\Delta_c)(\Delta_Z^2+\Delta_c^2)}\biggr)^{-1/2}.
\end{aligned}
\end{equation}
Substituting Eq.~\eqref{normalization} into Eq.~\eqref{WFs}, we see that the two Majorana wave functions are orthogonal,
\begin{equation}
\phi_2^\dagger(x)\phi_1(x)=0,
\end{equation}
and related by the emergent time-reversal symmetry ($\mathcal{T}_\od=i\tau_0\eta_x\sigma_y$),
\begin{equation}
\begin{aligned}
\phi_1(x)&=\mathcal{T}_\od\phi_2^*(x), \\
\phi_2(x)&=-\mathcal{T}_\od\phi_1^*(x).
\end{aligned}
\end{equation}
Hence, the two Majorana bound states form a Kramers pair that is protected by the emergent time-reversal symmetry.

Constructing a general zero-energy solution to Eq.~\eqref{BdG} in the trivial phase ($\Delta_Z^2<\Delta_d^2-\Delta_c^2$) requires replacing
\begin{equation}
\begin{gathered}
k_{7,8}\to i\bigl(\Delta_d-\sqrt{\Delta_Z^2+\Delta_c^2}\bigr)/\alpha, \\
\chi_7\to\frac{1}{2\sqrt{\Delta_Z^2+\beta^2}}(-\beta,-i\Delta_Z,\beta,-i\Delta_Z,-i\beta,\Delta_Z,i\beta,\Delta_Z)^T \\
\chi_8\to \frac{1}{2\sqrt{\Delta_Z^2+\beta^2}}(i\Delta_Z,-\beta,i\Delta_Z,\beta,\Delta_Z,i\beta,\Delta_Z,-i\beta)^T.
\end{gathered}
\end{equation}
After making these replacements in Eq.~\eqref{WFexpansion}, we find that the boundary condition $\phi(0)=0$ can only be satisfied by choosing $c_{1-8}=0$ and hence there are no Majorana bound states.

\newpage
\end{widetext}

\end{document}